\documentclass[aps,pre,twocolumn,superscriptaddress,showpacs]{revtex4-1}
\usepackage{amsmath,amssymb,graphicx,color}
\usepackage{bm}
\usepackage{bbm}
\usepackage{mathrsfs}
\def\Tr{\mbox{Tr}}

\begin{document}
\title{Quantum fluctuation theorems and generalized measurements during the force protocol}

\author{Gentaro Watanabe}
\affiliation{Asia Pacific Center for Theoretical Physics (APCTP), San 31, Hyoja-dong, Nam-gu, Pohang, Gyeongbuk 790-784, Korea}
\affiliation{Department of Physics, POSTECH, San 31, Hyoja-dong, Nam-gu, Pohang, Gyeongbuk 790-784, Korea}
\author{B. Prasanna Venkatesh}
\affiliation{Asia Pacific Center for Theoretical Physics (APCTP), San 31, Hyoja-dong, Nam-gu, Pohang, Gyeongbuk 790-784, Korea}
\author{Peter Talkner}
\affiliation{Institut f\"{u}r Physik, Universit\"{a}t Augsburg, Universit\"{a}tsstra\ss e 1, D-86135 Augsburg, Germany}
\affiliation{Asia Pacific Center for Theoretical Physics (APCTP), San 31, Hyoja-dong, Nam-gu, Pohang, Gyeongbuk 790-784, Korea}
\author{Michele Campisi}
\affiliation{Institut f\"{u}r Physik, Universit\"{a}t Augsburg, Universit\"{a}tsstra\ss e 1, D-86135 Augsburg, Germany}
\author{Peter H\"anggi}
\affiliation{Institut f\"{u}r Physik, Universit\"{a}t Augsburg, Universit\"{a}tsstra\ss e 1, D-86135 Augsburg, Germany}
\affiliation{Nanosystems Initiative Munich, Schellingstra\ss e 4, D-80799 M\"unchen, Germany}
\date{\today}
\begin{abstract}
Generalized measurements of an observable performed on a quantum system during a force protocol are investigated and conditions that guarantee the validity of the Jarzynski equality and the Crooks relation are formulated. 
In agreement with previous studies by Campisi {\it et al.} [M. Campisi, P. Talkner, and P. H\"anggi, Phys. Rev. Lett. {\bf 105}, 140601 (2010); Phys. Rev. E {\bf 83}, 041114 (2011)], we find that these fluctuation relations are satisfied for projective measurements; however, for generalized measurements special conditions on the operators determining the measurements need to be met. For the Jarzynski equality to hold, the measurement operators of the forward protocol must be normalized in a particular way. 
The Crooks relation
additionally entails that the backward and forward measurement operators depend on each other. Yet, quite some freedom is left as to how the two sets of operators 
are interrelated.
This ambiguity is removed if one considers selective measurements, which are specified by a {\it joint} probability density function of work and measurement results of the considered observable. We find that the respective forward and backward joint probabilities satisfy the Crooks relation only if the measurement operators of the forward and backward protocols are the time-reversed adjoints of each other. In this case, the work probability density function {\it conditioned} on the measurement result satisfies a modified Crooks relation. The modification appears as a protocol-dependent factor that can be expressed by the 
information gained by the measurements during the forward and backward protocols.
Finally, detailed fluctuation theorems with an arbitrary number of intervening measurements are obtained.
\end{abstract}
\maketitle

\section{Introduction}

The fluctuation theorems of Jarzynski \cite{J} and Crooks \cite{C}, with their precursors by Bochkov and Kuzovlev \cite{BK1,BK2}, have proved very robust, being valid in a very wide range of situations. Originally formulated for classical closed systems, their validity  was subsequently demonstrated for open classical \cite{J04}, as well as for closed \cite{K,T,TLH} and open quantum systems \cite{TCH09,CTH09}. References \cite{EHM,CHT11} provide recent reviews on quantum fluctuation theorems.

The Crooks and Jarzynski fluctuation theorems relate the statistics of work performed on a system by a force with prescribed time dependence and of finite duration to an isothermal change of free energy.
For quantum systems, the validity of these transient fluctuation theorems rests on two assumptions: (i) The considered system initially stays in thermal equilibrium described by a canonical density matrix \cite{fn1} and (ii) the system undergoes Hamiltonian evolution and hence time-reversible dynamics. Under these assumptions, the fluctuation theorems hold also for open systems provided the dynamics of the environment is taken into account on the level of Hamiltonian dynamics \cite{J04,CTH09,fn2}.

For quantum systems, determining the work requires two measurements of energy, one immediately before the force starts to act and the second one at the end of the force protocol \cite{K,TLH}. These measurements typically are treated as projective measurements \cite{WM}; only recently, the influence of generalized measurements was investigated \cite{VWT}. The main result of Ref.~\cite{VWT} is that so-called universal energy measurements --- these are measurements that are not particularly adapted to an actual force protocol --- must be projective in order that the work statistics complies with the fluctuation theorems.

Measurements of observables other than energy during the force protocol have been considered in the context of feedback control. The result of the measurement of a control variable can be used to adjust the protocol leading to a modification of the Jarzynski equality and the Crooks relation \cite{SU,HV,MT}. Related topics  occur in the contexts of Maxwell demons, Szilard engines, and Landauer's erasure principle \cite{MNV,Sag,TSUMS,S,L,VJ1,DL}.

It was pointed out that projective nonselective measurements during the force protocol do not alter these fluctuation relations \cite{CTH10,CTH11}, even though the statistics of work may be drastically changed. As it was already noticed in Ref.~\cite{MT}, the fluctuation relations do not always hold for generalized measurements. In the present paper we investigate the conditions under which the fluctuation theorems stay valid if generalized measurements intervene a force protocol. These measurements are supposed to be universal, which means, as already mentioned, that the way the measurements are performed does not depend on the force protocol.

After a short review of the formalism describing generalized measurements in Sec.~\ref{GM}, we characterize the statistics of work in presence of intervening measurements in Sec.~\ref{WS}. In Sec.~\ref{FT} we investigate the consequences of the fluctuation theorems on the measurement operators for nonselective intervening measurements.  Selective intervening measurements are discussed in Sec.~\ref{SM}. With a proper choice of the measurement operators, the Crooks relation is obtained for the joint probability density function (PDF) of work and measured observable. For the same choice of measurement operators, only a modified Crooks relation holds if the work pdf is conditioned on the outcome of the intervening measurement.
For an arbitrary number of intervening measurements, detailed fluctuation theorems are obtained.   
The paper closes with a discussion in Sec.~\ref{Con}. Technical details concerning the conditions on the measurement operators are presented in Appendices~\ref{A1} and \ref{A2}.

\section{Generalized measurements}
\label{GM}
We collect here some notions of the theory of generalized measurements that we will need later on. For a more complete presentation, we refer to the literature, e.g., Ref.~\cite{WM}.

The generalized measurement of an observable $A$ with spectral representation $A = \sum_i a_i \Pi^A_i$, expressed in terms of  the eigenvalues $a_i$ and the complete set of orthogonal eigenprojection operators $\Pi^A_i$, is formally characterized by  measurement operators $M_x$.  Here $x$ denotes the position of a pointer, which may assume values in a range $X$. This range of pointer positions may coincide with the discrete set of indices labeling the eigenvalues and eigenprojectors of the observable $A$.  
Alternatively, the pointer set $X$ may be given by the range of the classically equivalent observable $A^{\text{classical}}$. Then $X$ is continuous and contains all eigenvalues $a_i$ as isolated points. 
In either case, when a measurement has a particular value $x$ as a result, the state of the system immediately after this selective measurement, also known as a conditional measurement in quantum optics, is given by the density matrix
\begin{equation}
\rho_x = M_x \rho M^\dagger_x/p_x(\rho)\:,
\label{rx}
\end{equation}
where $\rho$ denotes the state immediately before the measurement. Here   
the probability  of obtaining the result $x$ in the state $\rho$, $p_x(\rho)$,  is given by
\begin{equation}
p_x(\rho) = \Tr\, M^\dagger_x M_x \rho\:.
\label{px}
\end{equation}

In the case where the measurement is performed and the outcome is not recorded, in other words, for a nonselective measurement, 
the state after the measurement is given by the integral over $X$ weighted by the probabilities $p_x(\rho)$ yielding the postmeasurement state
\begin{equation}
\rho^{\text{pm}} = \int_X dx\, M_x \rho M^\dagger_x\:.
\label{rpm}
\end{equation}
It is a distinguishing feature of quantum mechanics that a nonselective postmeasurement state differs in general from the state immediately before the measurement.
The normalization of the postmeasurement states $\Tr\, \rho^{\text{pm}} =1$ must be guaranteed for all density matrices $\rho$ prior to the measurement. This requirement implies a normalization condition for the measurement operators reading
\begin{equation}
\int_X dx\, M^\dagger_x M_x =   \mathbbm{1}\:,
\label{MM1}
\end{equation}
where $\mathbbm{1}$ denotes the unit operator on the Hilbert space of the considered system.
In the case of a discrete set $X$, the integrals in Eqs.~(\ref{rpm}) and (\ref{MM1})  have to be replaced by sums $\int_X dx \rightarrow \sum_{x\in X}$.
For a projective or von Neumann measurement of the observable $A$, the pointer positions are discrete and the measurement operators coincide with the eigenprojection operators of $A$: $M_i = \Pi^A_i$.

\section{work statistics with intervening generalized measurements}\label{WS}
We consider the action of an external time-dependent force $\lambda(t)$ on a system. The force $\lambda(t)$ varies in time according to a fixed protocol $\Lambda=\{\lambda(t)|0\leq t \leq \tau \}$ of total duration $\tau$. The system Hamiltonian depends on this parameter and is denoted by $H(\lambda(t))$. It is supposed to represent the energy of the system. 
{\it Projective measurements} of the energy at the {\it beginning} and the {\it end} of the protocol are performed in order to determine the work done on the system.
The time evolution, which is governed by the Hamiltonian  $H(\lambda(t))$, is interrupted at time $t_1 \in (0, \tau)$ by a generalized measurement of an observable $A$ by means of the measurement operators $M_x$, $x \in X$. The joint probability $p_\Lambda(m,\tau;x,t_1;n,0)$ to find the eigenvalue $e_n(0)$ of the Hamiltonian $H(\lambda(0))$ in the first energy measurement, the pointer position $x$ in the measurement of $A$ at time $t_1$, and the eigenvalue $e_m(\tau)$ of the Hamiltonian $H(\lambda(\tau))$ in the final energy measurement is given by 
\begin{equation}
\begin{split}
 p_\Lambda(m,\tau;x,t_1;n,0) &= \Tr\, \Pi_m(\tau) U_{\tau,t_1}(\Lambda)  M_x U_{t_1,0}(\Lambda) \\ & \quad \times \Pi_n(0)   \rho(0) 
\Pi_n(0)U^\dagger_{t_1,0}(\Lambda) M^\dagger_x\\ & \quad \times U^\dagger_{\tau,t_1}(\Lambda) \Pi_m(\tau)\:.
\end{split}
\label{pmxn}
\end{equation}
Here $\Pi_n(t) \equiv \Pi^{H(\lambda(t))}_n$ with $t=0,\, \tau$ is the projection operator onto the eigenspace of $H(\lambda(t))$ corresponding to the eigenenergy $e_n(t)$, and  $U_{t,s}(\Lambda)$ is the unitary time evolution operator which follows as the solution of the Schr\"odinger equation
\begin{equation}
\begin{split}
i \hbar \partial U_{t,s}(\Lambda) /\partial{t}& = H(\lambda(t))  U_{t,s}(\Lambda)\:,\\ 
U_{s,s}(\Lambda) &= \mathbbm{1}\:.
\label{UH}
\end{split}
\end{equation}
Further, $\rho(0)$ denotes the initial density matrix which is assumed to be given by the Gibbs state at inverse temperature $\beta$ reading
\begin{equation}
\rho(0) = Z^{-1}(0) e^{-\beta H(\lambda(0))}\:,
\label{Gibbs}
\end{equation}
with the partition function 
\begin{equation}
Z(0) = \Tr\, e^{-\beta H(\lambda(0))}\:.
\label{Z0}
\end{equation}

The PDF of work performed on the system, $p_\Lambda(w)$, follows as
\begin{equation}
\begin{split}
p_\Lambda(w) &= \sum_{m,n} \delta (w-e_m(\tau)+e_n(0))\\
&\quad \times \int_X dx\, p_\Lambda (m,\tau;x,t_1;n,0)\:.
\label{pw}
\end{split}
\end{equation}
Then the characteristic function of work, which is defined as the Fourier transform of the work PDF $G_\Lambda(u) = \int dw\, e^{iuw} p_\Lambda(w)$, can be expressed as
\begin{equation}
\begin{split}
G_\Lambda(u) &= \int_X dx\, \Tr\, U^\dagger_{t_1,0}(\Lambda)M^\dagger_x U^\dagger_{\tau,t_1}(\Lambda) e^{i uH(\lambda(\tau))}\\
&\quad \times U_{\tau,t_1}(\Lambda)
M_x U_{t_1,0}(\Lambda) e^{-iu H(\lambda(0))} \rho(0)\:.
\label{Gu}
\end{split}
\end{equation}
Generalization to multiple measurements during the force protocol 
will be considered below.
For projective measurements, the characteristic function (\ref{Gu}) agrees with the expression resulting from Eqs. (17) and (19) of Ref. \cite{CTH11}.
\section{Fluctuation theorems}\label{FT}
We now postulate the validity of the fluctuation theorems by Crooks and Jarzynski in the presence of a measurement at an instant of time during the force protocol, and study the consequent restrictions that apply for the measurement operators. We require that the measurement operators $M_x$ are {\it universal} in the sense that the validity of the fluctuation theorems is not restricted to particular protocols, but that they hold for all possible force protocols connecting any initial and final Hamiltonians.
We first consider the  Jarzynski equality, which is less restrictive.

\subsection{Jarzynski equality}\label{JEQ}

The Jarzynski equality 
\begin{equation}
\langle e^{-\beta w}\rangle = e^{- \beta \Delta F}
\label{JE}
\end{equation}
relates the expectation of the exponentiated work to the difference between free energies of the system for the initial and final parameter values, both in thermal equilibrium at the initial temperature.  It can equivalently be expressed in terms of the characteristic function of work as
\begin{equation}
Z(0) G_\Lambda(i \beta)/Z(\tau) = 1\:,
\label{JG}
\end{equation}
with
\begin{equation}
Z(\tau) = \Tr\, e^{-\beta H(\lambda(\tau))}\:.
\label{Zt}
\end{equation}
The free energy difference is given by $\Delta F = - \beta^{-1} \ln [Z(\tau)/Z(0)]$.  
Putting the explicit form (\ref{Gu}) into the left-hand side of Eq.~(\ref{JG}),
we find
\begin{equation}
\begin{split}
Z(0) G_\Lambda(i \beta)/Z(\tau) &= Z^{-1}(\tau) \Tr \int_X dx\, M_x M^\dagger_x\\ &\quad \times U^\dagger_{\tau,t_1}(\Lambda) e^{-\beta H(\lambda(\tau))} U_{\tau,t_1}(\Lambda)\:.
\label{CJE}
\end{split}
\end{equation}  
The right-hand side of this equation becomes unity and hence the Jarzynski equality is satisfied if the integral of the product of the measurement operator and its adjoint gives unity, i.e., if 
\begin{equation}
\int_X dx\, M_x M^\dagger_x = \mathbbm{1}
\label{MMt1} 
\end{equation}
holds \cite{MT}. Note that the resolution of unity in terms of the operators $M^\dagger_xM_x$ [Eq.~(\ref{MM1})] does not imply the respective relation (\ref{MMt1}) in terms of $M_x M^\dagger_x$. It does so for normal measurement operators and in particular for self-adjoint measurement operators, which are called {\it minimally disturbing} according to Wiseman and Milburn \cite{WM}.  Projective measurements fall into this class.

The resolution of unity in terms of $M_x M^\dagger_x$, given by Eq.~(\ref{MMt1}), is not only a sufficient but also a necessary condition in order that the Jarzynski equality holds for universal measurements. The main step in proving the necessity is based on the fact that, for universal measurements, the time evolution operator $U_{\tau,0}(\Lambda)$ and consequently also  $U_{\tau, t_1}(\Lambda)$ can be chosen arbitrarily \cite{VWT}. Further details of the proof are presented in  Appendix~\ref{A1}.

In passing, we note that a set of operators ${M_x}$ that fulfills both relations (\ref{MM1}) and (\ref{MMt1}) defines a unital map  $\mathcal{M}\rho = \int_X dx\, M_x \rho M^\dagger_x$ acting on density matrices.

\subsection{Crooks relation}\label{CRE}

The Crooks relation connects the statistics of the forward protocol $\Lambda=\{ \lambda(t)|0\leq t \leq \tau \}$ with  the work statistics of the backward protocol $\bar{\Lambda} = \{ \epsilon_\lambda \lambda(\tau -t)|0\leq t \leq \tau \}$
according to which $\lambda(t)$ is replaced by the time-reversed force parameter $\epsilon_\lambda \lambda(\tau-t)$, where $\epsilon_\lambda$ is the parity of $\lambda$ under time reversal, such as $\epsilon_E =1$ for an electric field or $\epsilon_H = -1$ for a magnetic field. Further, the force parameter is run through in the reverse order. The initial state of the backward protocol is the canonical  equilibrium state at the inverse temperature $\beta$ and at the parameter value $\epsilon_\lambda \lambda(\tau)$.  
The Crooks relation can be expressed in terms of the work PDFs $p_\Lambda(w)$ and $p_{\bar{\Lambda}}(w)$ as
\begin{equation}
p_\Lambda(w) = e^{-\beta (\Delta F -w)} p_{\bar{\Lambda}}(-w)\:,
\label{CRp}
\end{equation}
or, equivalently in terms of the respective characteristic functions $G_\Lambda(u)$ and $G_{\bar{\Lambda}}(u)$ as
\begin{equation}
Z(0) G_\Lambda(u) = Z(\tau) G_{\bar{\Lambda}}(-u +i\beta)
\label{CRG}
\end{equation}
(see Refs.~\cite{C,T,TH}).
The Crooks relation follows from the initial canonical equilibrium and from the reversibility of systems with time-dependent Hamiltonian obeying instantaneous time reversibility of the form 
\begin{equation}
\theta H(\lambda(t)) \theta^\dagger = H(\epsilon _\lambda \lambda(t))\:,
\label{Hth}
\end{equation}
which relates the inverse time-evolution operator of the forward process to the time-evolution operator of the time-reversed process \cite{AG,CHT11}
\begin{equation}
U^{-1}_{t,s}(\Lambda) =U^\dagger_{t,s}(\Lambda) = \theta^\dagger U_{\tau-s,\tau-t}(\bar{\Lambda}) \theta\:. 
\label{Uth}
\end{equation}
Here $\theta$ denotes the antiunitary time-reversal operator \cite{M}.

We now investigate the question under which conditions the Crooks relation continues to hold in the presence of an intermediate measurement of an observable by means of the measurement operator $M_x$ at time $t_1$. For this purpose, we postulate the validity of the Crooks relation for the characteristic function of the forward protocol as given by Eq.~(\ref{Gu}) and the corresponding characteristic function for the backward protocol, which is given by
\begin{equation}
\begin{split}
G_{\bar{\Lambda}}(u) &= \int_X dx\, \Tr\, U^\dagger_{\tau-t_1,0}( \bar{\Lambda}) \tilde{M}^\dagger_x U^\dagger_{\tau,\tau-t_1}( \bar{\Lambda})\\
&\quad \times  
e^{iuH(\epsilon_\lambda \lambda(0))} U_{\tau,\tau-t_1}( \bar{\Lambda}) \tilde{M}_x U_{\tau-t_1,0}( \bar{\Lambda})\\
&\quad \times   e^{-iu H(\epsilon_\lambda\lambda(\tau))}\bar{\rho}(\tau)\:.
\label{Gubp}
\end{split}
\end{equation} 
Here $\tilde{M}_x$ denotes the operator describing the measurement of the time-reversed observable $\bar{A} = \theta A \theta^\dagger$ during the backward protocol at time $\tau-t_1$. The precise form of $\tilde{M}_x$ is still left open. The initial condition of the backward process is given by the time-reversed canonical state at the final parameter value, i.e., by $\bar{\rho}(\tau) = Z^{-1}(\tau) e^{-\beta H(\epsilon_\lambda \lambda(\tau))}$.
Putting the explicit forms (\ref{Gu}) and (\ref{Gubp}) of the characteristic functions into the Crooks relation (\ref{CRG}) and using the time-reversal symmetry (\ref{Uth}), one obtains the condition 
\begin{equation}
\begin{split}
&\int_X dx\, \Tr\, U^\dagger_{t_1,0} (\Lambda)\bar{\tilde{M}}_x U^\dagger_{\tau,t_1}(\Lambda) e^{iu H(\lambda(\tau))} \\
& \quad \times U_{\tau,t_1} (\Lambda)  \bar{\tilde{M}}^\dagger_x U_{t_1,0} (\Lambda) e^{-iu H(\lambda(0))} e^{-\beta H(\lambda(0))}\\
&=  \int_X dx\, \Tr\, U^\dagger_{t_1,0} (\Lambda) M^\dagger_x U^\dagger_{\tau,t_1}(\Lambda) e^{iu H(\lambda(\tau))}\\
& \quad \times  U_{\tau,t_1} (\Lambda)  M_x U_{t_1,0} (\Lambda) e^{-iu H(\lambda(0))} e^{-\beta H(\lambda(0))}\:,
\end{split}
\label{CRc1}
\end{equation}
with a bar denoting time reversal, i.e., 
\begin{equation}
\bar{\tilde{M}}_x = \theta^\dagger \tilde{M}_x \theta\:.
\label{bar}
\end{equation}
Obviously, the condition (\ref{CRc1}) is satisfied and consequently the Crooks relation holds if the measurement operators in the backward protocol are chosen as the time-reversed adjoint measurement operators of the forward protocol, i.e., if
\begin{equation}
\tilde{M}_x = \theta M^\dagger_x \theta^\dagger
\label{MthM}
\end{equation}  
holds.
This also presupposes that  $M_x M^\dagger_x$ must add up to the identity in order that $\tilde{M}_x$ is a properly normalized measurement operator. Hence we recover the necessary and sufficient condition for the Jarzynski equality (\ref{MMt1}).

Equation (\ref{MthM}) presents a particular choice of the measurement operators that may be employed in the backward process such that the Crooks relation is satisfied.
However, there exist other measurement operators that also yield the Crooks relation. 
All backward measurement operators that depend on the forward measurement operators via an integral transform
\begin{equation} 
\bar{\tilde{M}}_x = \int_X dy\, f(x,y) M^\dagger_y
\label{MfM}
\end{equation}
with a complex-valued integral kernel $f(x,y)$ satisfying
\begin{equation}
\int_X dx\, f(x,y) f^*(x,z) = \delta(y-z)
\label{fff}
\end{equation}
satisfy Eq.~(\ref{CRc1}). It seems plausible that there are no other solutions of Eq.~(\ref{CRc1}) apart from these linear combinations, even though a formal proof of this conjecture is missing.

In the following section, we will find a more definite result in the case of selective measurements.

\section{Selective measurements}\label{SM}

In a selective measurement during a force protocol, the pointer position $x$ is registered together with the energies at the beginning and the end of the protocol. The result is characterized by a joint PDF of work and pointer position $p_\Lambda(w,x)$, which is given by
\begin{equation}
\begin{split}
p_\Lambda(w,x)& = \sum_{m,n} \delta( w- e_m(\tau) + e_n(0))\\
&\quad \times p_\Lambda(m,\tau;x,t_1;n,0)\:,      
\label{pwx}
\end{split}
\end{equation}
with the joint PDF $ p_\Lambda(m,\tau;x,t_1;n,0)$ being defined by Eq.~(\ref{pmxn}).
Its marginal yields the work PDF (\ref{pw}) for a nonselective measurement
\begin{equation}
p_\Lambda(w) = \int_X dx\, p_\Lambda(w,x)\:.
\label{pwpwx}
\end{equation}
Accordingly, the characteristic function of work for the selective measurement
\begin{equation}
\begin{split}
G_\Lambda(u,x)& = \Tr\, U^\dagger_{t_1,0}(\Lambda) M^\dagger_x U^\dagger_{\tau,t_1}(\Lambda)e^{iuH(\lambda(\tau))}\\
&\quad \times  U_{\tau,t_1}(\Lambda)  M_x U_{t_1,0}(\Lambda)  e^{-iuH(\lambda(0))} \rho(0)
\label{Gux}
\end{split}
\end{equation}
gives the non-selective measurement characteristic function (\ref{Gu}) upon integration over $x$  
\begin{equation}
G_\Lambda(u) = \int_X dx\, G_\Lambda(u,x)\:.
\label{GuGux}
\end{equation}

Using analogous arguments as for the non-selective case, one obtains for the selective case a Crooks relation
of the form
\begin{equation}
Z(0) G_\Lambda(u,x) = Z(\tau) G_{\bar{\Lambda}}(-u +i\beta,x)
\label{CRGx}
\end{equation}
if and only if the integrands of the left- and right-hand sides of Eq.~(\ref{CRc1})
agree with each other, i.e., if
\begin{equation}
\begin{split}
& \Tr\, U^\dagger_{t_1,0} (\Lambda)\bar{\tilde{M}}_x U^\dagger_{\tau,t_1}(\Lambda) e^{iu H(\lambda(\tau))} \\
&  \times U_{\tau,t_1} (\Lambda) \bar{\tilde{M}}^\dagger_x U_{t_1,0}(\Lambda) e^{-iu H(\lambda(0))} e^{-\beta H(\lambda(0))}\\
&=  \Tr\, U^\dagger_{t_1,0} (\Lambda) M^\dagger_x U^\dagger_{\tau,t_1}(\Lambda) e^{iu H(\lambda(\tau))}\\
& \times  U_{\tau,t_1} (\Lambda) M_x U_{t_1,0}(\Lambda) e^{-iu H(\lambda(0))} e^{-\beta H(\lambda(0))}
\end{split}
\label{CRxc1}
\end{equation}
holds. This condition is fulfilled if and only if the measurement operators of the forward and backward protocols are related by Eq.~(\ref{MthM}).
The sufficiency of this condition is obvious. Its necessity is demonstrated in Appendix~\ref{A2}.

The Crooks relation for selective measurements can equivalently be expressed in terms of the joint PDF
as
\begin{equation}
p_\Lambda(w,x) = e^{-\beta (\Delta F -w)} p_{\bar{\Lambda}}(-w,x)\:.
\label{CRpx}
\end{equation}
This implies a modified Jarzynski equality for the conditional exponential expectation of work reading
\begin{equation}
\langle e^{-\beta w} \rangle_x = e^{-\beta \Delta F}\frac{p_{\bar{\Lambda}}(x)}{p_\Lambda(x)}\:,
\label{JEx}
\end{equation}
where the conditional average $\langle \cdot \rangle_x$ is taken with respect to the conditional probability 
\begin{equation}
p_\Lambda(w|x) = p_\Lambda(w,x)/ p_\Lambda(x)\:,
\label{pcwx}
\end{equation}
with the marginal forward PDF of $x$ defined by
\begin{equation}
\begin{split}
 p_\Lambda(x)&=\int dw\, p_\Lambda(w,x)\\
&= \Tr\, M^\dagger_x M_x U_{t_1,0}(\Lambda) \rho(0) U^\dagger_{t_1,0}(\Lambda)\:.
\end{split}
\label{pfx1} 
\end{equation}
Accordingly, the marginal backward PDF of $x$ is given by
\begin{equation}
\begin{split}
 p_{\bar{\Lambda}}(x)&=\int dw\, p_{\bar{\Lambda}}(w,x)\\
&= \Tr\, M_x M^\dagger_x U^\dagger_{\tau,t_1}(\Lambda) \rho(\tau) U_{\tau,t_1}(\Lambda)\:.
\end{split}
\label{pbx1} 
\end{equation} 
Note that the modification of the Jarzynski equality caused by selective measurements during the force protocol is determined by a protocol-dependent correction factor. This factor can be expressed in terms of the difference in the entropies of the backward and forward marginal $x$ PDFs, which are given by $I_\Lambda(x) = - \ln p_\Lambda(x)$ and $I_{\bar{\Lambda}}(x) = - \ln p_{\bar{\Lambda}}(x)$, respectively. Hence, the Jarzynski equality can be written as
\begin{equation}
\langle e^{-\beta w} \rangle_x = e^{-[\beta \Delta F +\Delta I(x)]}\:,
\label{JEcx}
\end{equation}  
where $\Delta I(x) = I_{\bar{\Lambda}}(x) -I_\Lambda(x)$  denotes the entropy difference, i.e., the difference in the information gains in measurements of $x$ during the forward and backward protocols.
The same difference of the information gains also appears as a correction factor in the Crooks relation for the conditional work PDF, reading
\begin{equation}
p_\Lambda(w|x) = e^{-\beta[\Delta F + \beta^{-1} \Delta I(x) - w]}\, p_{\bar{\Lambda}}(-w|x)\:. 
\label{CRcx}
\end{equation}

The same expression for the conditional average of the exponentiated work as in Eq.~(\ref{JEx}) was obtained for a classical process in Ref.~\cite{KPVdB}. 
For classical processes subject to feedback control a modified Jarzynski equality of the form of Eq.~(\ref{JEcx}) was derived in Ref.~\cite{SU}, 
with the difference that the  protocol depends on the controlled observable, i.e., on its measured value. 
The quantum version of the Jarzynski equality in the presence of feedback control was derived in Ref.~\cite{MT}.
A modified Crooks relation of the form of Eq.~(\ref{CRcx}) was obtained in Ref.~\cite{HV} for classical feedback control.

\subsection{Detailed fluctuation theorem}

Finally we consider relations between the forward and backward joint distributions specifying the likelihood of finding energy values $e_n(0)$ and $e_m(\tau)$, in combination with the results $x_i$ of $k$ measurements of an observable $A$ at consecutive times $0<t_1< \cdots <t_i< \cdots <t_k < \tau$. 
We denote the sequence of the pairs consisting of the measurement results and the respective instants of these measurements by $\mathbbm{X}_k=  \{ x_k,t_k; x_{k-1},t_{k-1}; \ldots; x_1, t_1 \}$ during the forward protocol and by $\bar{\mathbbm{X}}_k=  \{ x_1,\tau-t_1; x_2,\tau-t_2; \ldots; x_k, \tau-t_k \}$ during the backward protocol.
For each measurement, the same measurement operators $M_x$, $x\in X$ satisfying Eqs. (\ref{MM1}) and (\ref{MMt1}) are used in the forward protocol. The time-reversed adjoint measurement operators are employed in the backward protocol such that, in the case of a single intervening measurement, the Crooks relation is guaranteed to hold.      
The joint distribution of two energy measurements and $k$ $A$-measurements, $p_\Lambda(m,\tau; \mathbbm{X}_k;n,0)$, is then given by
\begin{equation}
\begin{split}
&p_\Lambda(m,\tau; \mathbbm{X}_k;n,0)  =\Tr\, \Pi_m(\tau) U_{\tau,t_k}(\Lambda) M_{x_k}\\
&\times U_{t_k,t_{k-1}}(\Lambda) M_{x_{k-1}}  U_{t_{k-1},t_{k-2}}(\Lambda) \times \cdots \times U_{t_2,t_1}(\Lambda) M_{x_1} \\
&\times U_{t_1,0}(\Lambda) \Pi_n(0) \rho(0) \Pi_n(0) U^\dagger_{t_1,0}(\Lambda)  M^\dagger_{x_1} U^\dagger_{t_2,t_1}(\Lambda)  \times \cdots \\
& \times U^\dagger_{t_{k-1},t_{k-2}}(\Lambda) 
M^\dagger_{x_{k-1}} U^\dagger_{t_k,t_{k-1}}(\Lambda) M^\dagger_{x_k} U^\dagger_{\tau,t_k}(\Lambda) 
\end{split}
\label{pmXnf}
\end{equation}
for the forward protocol. For a single intervening measurement $k=1$, it coincides with the expression (\ref{pmxn}). 
Analogous to Eq.~(\ref{pmXnf}), 
one obtains the joint distribution of energies and intervening measurements for the backward protocol. It reads
\begin{equation}
\begin{split}
&p_{\bar{\Lambda}}(n,0; \bar{\mathbbm{X}}_k;m,\tau) = \Tr\, \bar {\Pi}_n(0) U_{\tau, \tau-t_1} (\bar{\Lambda}) \tilde{M}_{x_1} \\ 
& \times U_{\tau-t_1, \tau-t_2}(\bar{\Lambda})
 \tilde{M}_{x_2} U_{\tau-t_2, \tau-t_3}(\bar{\Lambda}) \\
&\times  \cdots \times 
U_{\tau-t_{k-1},\tau-t_k} (\bar{\Lambda}) \tilde{M}_{x_k}U_{\tau-t_k,0}(\bar{\Lambda}) \bar{\Pi}_m(\tau) \bar{\rho}(\tau) \bar{\Pi}_m(\tau)\\ 
& \times U^\dagger_{\tau-t_k,0}(\bar{\Lambda})\tilde{M}^\dagger_{x_k} U^\dagger_{\tau-t_{k-1},\tau-t_k} (\bar{\Lambda}) \times  \cdots \times U^\dagger_{\tau-t_2, \tau-t_3}(\bar{\Lambda}) \\
& \times \tilde{M}^\dagger_{x_2} U^\dagger_{\tau-t_1, \tau-t_2}(\bar{\Lambda}) \tilde{M}^\dagger_{x_1} U^\dagger_{\tau, \tau-t_1} (\bar{\Lambda})\:.
\end{split}
\label{pnXmb}
\end{equation}
Applying the time-reversal relation (\ref{Uth}) and replacing the backward by the forward  measuring operators by means of Eq.~(\ref{MthM}), one finds that these two distributions are proportional to each other. Their ratio obeys the detailed fluctuation theorem
\begin{equation}
\frac{p_\Lambda(m,\tau; \mathbbm{X}_k; n,0)}{p_{\bar{\Lambda}}(n,0;\bar{\mathbbm{X}}_k;m,\tau)} = e^{-\beta [\Delta F -e_m(\tau) + e_n(0)]}\:,
\label{dft}
\end{equation}  
which is of the same form as it holds for projective measurements of $A$ \cite{CHT11} and for classical Hamiltonian dynamics \cite{BK1}.

By taking the logarithm on both sides of Eq.~(\ref{dft}) and averaging over all initial and final states, as well as over all intervening measurement results, one finds the Kullback-Leibler divergence of the backward distribution from the forward distribution to agree with the irreversible work in the presence of $k$ intervening nonselective measurements of an observable $A$, i.e.,
\begin{equation}
\beta \left (\langle w \rangle^{A}_{T_k} - \Delta F \right ) = D(p_\Lambda|| p_{\bar{\Lambda}})\:,
\label{KLDW}
\end{equation}
where the irreversible work is the difference between the average work $\langle w \rangle^{A}_{T_k}$ and the free energy change $\Delta F$. The average work is obtained as
\begin{equation}
\begin{split}
\langle w \rangle^A_{T_k}& = \int_{X^k} dx_1\cdots dx_k \sum_{m,n} [e_m(\tau)-e_n(0)] \\
& \quad \times p_\Lambda(m,\tau;\mathbbm{X}_k;n,0)\:.
\end{split}
\label{wAT}
\end{equation} 
Note that, in contrast to a classical process, the 
average work depends on the $k$-fold $A$-measurements as indicated by the superscript $A$ and on the series of instants of measurements $T_k = (t_1,\ldots, t_k)$
even though integrations over all results of these measurements are performed.

The Kullback-Leibler divergence is defined as
\begin{equation}
\begin{split}
D(p_\Lambda||p_{\bar{\Lambda}}) & = \int_{X^k} dx_1\cdots dx_k \sum_{m,n}p_\Lambda(m,\tau;\mathbbm{X}_k;n,0)\\
&\quad \times \ln \frac{p_\Lambda(m,\tau;\mathbbm{X}_k;n,0)}{p_{\bar{\Lambda}}(n,0;\bar{\mathbbm{X}}_k;m,\tau)}\:.
\end{split}
\label{KL}
\end{equation}
It is a measure of the distance between the distributions describing the forward and backward processes and has been related to the arrow of time \cite{FC,CH} and to dissipation \cite{KPVdB,VJ2}.

As for the joint work and single-measurement PDF, we can introduce the joint distribution of initial and final energies conditioned on a prescribed series $\mathbbm{X}_k$ of $k$ intervening measurements. In analogy to Eq.~(\ref{pcwx}), it reads
\begin{equation}
p_\Lambda(m,\tau;n,0|\mathbbm{X}_k) =\frac{p_\Lambda(m,\tau;\mathbbm{X}_k;n,0)}{p_\Lambda(\mathbbm{X}_k) }\:,
\label{pmnX}
\end{equation} 
where the marginal distribution of the measurement results is given by
\begin{equation}
p_\Lambda(\mathbbm{X}_k)= \sum_{m,n} p_\Lambda(m,\tau;\mathbbm{X}_k;n,0)\:.
\label{pX} 
\end{equation}
This leads to a modified detailed fluctuation theorem for the conditional distributions
\begin{equation}
\frac{p_\Lambda(m,\tau; n,0| \mathbbm{X}_k)}{p_{\bar{\Lambda}}(n,0;m,\tau|\bar{\mathbbm{X}}_k)} = e^{-\beta [\Delta F -e_m(\tau) + e_n(0)]}\, e^{-\Delta I(\mathbbm{X}_k)}\:,
\label{dftX}
\end{equation}
where the conditional distribution of the backward process,  $p_{\bar{\Lambda}}(n,0;m,\tau|\bar{\mathbbm{X}}_k)$, is defined analogously to the respective distribution of the forward process. 
The correction factor is determined by the difference 
between the information gain of the forward and backward processes
$\Delta I(\mathbbm{X}_k)=I_{\bar{\Lambda}} (\bar{\mathbbm{X}}_k) - I_{\Lambda} (\mathbbm{X}_k)$ due to the measurements in the forward and backward protocols, where $I_\Lambda(\mathbbm{X}_k) = - \ln p_\Lambda(\mathbbm{X}_k)$ and 
$I_{\bar{\Lambda}}(\bar{\mathbbm{X}}_k) = - \ln p_{\bar{\Lambda}}(\bar{\mathbbm{X}}_k)$.

The modified detailed fluctuation theorem is similar to the expression obtained in Ref.~\cite{HV} for a classical process with feedback control, with the main difference that, in contrast to a feedback controlled process, here the force protocol is not influenced by the results of the measurements.

\section{Conclusions}\label{Con}

We determined the statistics of work performed on a quantum system by a prescribed force protocol that is interrupted by a measurement of an observable $A$ and asked about the validity of the Jarzynski equality and the Crooks relation. Thereby we restricted ourselves to situations in which the work done on the system is determined by two projective measurements of the system's energy at the end and the beginning of the force protocol. Other, generalized initial and final energy measurements would typically lead to violations of the fluctuation theorems already in the absence of intermediate measurements \cite{VWT}. For projective intervening measurements of an observable $A$, the two fluctuation relations are known to hold \cite{CTH10,CTH11}.  

Generalized measurements are specified by measurement operators $M_x$ where $x$ is the pointer position, which indicates the measurement result. 
We found that the Jarzynski equality continues to hold if the map $\mathcal{M}u = \int_X dx\, M_x u M^\dagger_x$ can be defined on bounded operators $u$ and if it is unital, i.e., if $\mathcal{M} \mathbbm{1} = \mathbbm{1}$.
While for the Jarzynski equality this condition is necessary and sufficient, for the Crooks relation it is only necessary. 

The Crooks relation is satisfied if the measurement operators of the backward process are given by the time-reversed adjoint measurement operators of the forward process as expressed in Eq.~(\ref{MthM}). There are also other backward measurement operators, given by all possible normalization-conserving linear combinations,  for which the Crooks relation is satisfied when the measurement is nonselective.  

For selective measurements, we could demonstrate that Eq.~(\ref{MthM})  presents the only mutual assignment of the forward and backward measurement operators for which the Crooks relation is satisfied. However, the exponential work expectation conditioned on a measurement result  only conforms with a modified Jarzynski equality, containing  a correction factor that is determined by the difference of the information gains of the marginal backward and forward $x$ PDFs. 
The same protocol-dependent correction factor representing the difference in information gain due to the measurements during the forward and backward protocols also enters the Crooks relation for the conditional work PDF.   

We demonstrated that, under the same condition (\ref{MthM}) on the measurement operators in the forward and backward processes, the joint distributions of initial and final energies and of a set of intervening measurements obey detailed fluctuation relations. As in the case of classical closed systems (see, e.g., Ref.~\cite{HV}), the Kullback-Leibler divergence of the backward from the forward joint distributions agrees with the dissipated work.

\begin{acknowledgments} 
We acknowledge the Max Planck Society and the Korea Ministry of Education, Science and Technology (MEST), Gyeongsangbuk-Do, Pohang City, for the support of the JRG at APCTP. We were also supported by Basic Science Research Program through the National Research Foundation of Korea funded by the MEST (Grant No. 2012R1A1A2008028). P.H. and M.C. acknowledge support from the Volkswagen Foundation Project No. I/83902.
\end{acknowledgments}

\appendix
\section{Necessary condition for the Jarzynski equality}
\label{A1}

Combining Eqs. (\ref{JG}) and (\ref{CJE}), we obtain the relation
\begin{equation}
1 = Z^{-1}(\tau) \Tr \int_X dx\, M_x M^\dagger_x U^\dagger_{\tau,t_1}(\Lambda) e^{-\beta H(\lambda(\tau))} U_{\tau,t_1}(\Lambda)
\label{eJE}
\end{equation}
as an equivalent formulation of the Jarzynski equality. If the measurement operators are universal, this relation must hold for all protocols. This means that Eq.~(\ref{eJE}) must also be satisfied when the Hamiltonian is suddenly switched to any other Hamiltonian immediately after the beginning of the protocol. As a consequence, the condition (\ref{eJE}) must be valid for arbitrary unitary time-evolution operator $U$ replacing $U_{\tau,t_1}(\Lambda)$. Expressed by the spectral representation
\begin{equation}
 U = \sum_k e^{i \phi_k} |\varphi_k\rangle \langle \varphi_k|\:,
\label{U}
\end{equation}  
the resulting condition
\begin{equation}
1= \sum_{k,l} e^{i(\phi_k -\phi_l)} c_{k,l}
\label{eJE1}
\end{equation}
must hold for all possible values of the phases $\phi_k \in [0,2 \pi]$. The coefficients $c_{k,l}$ are defined as
\begin{equation}
c_{k,l}= \langle \varphi_k |\int_X dx\, M_x M^\dagger_x | \varphi_l \rangle \langle \varphi_l |\rho(\tau) |\varphi_k \rangle \:,
\label{ckl}
\end{equation}
where 
\begin{equation}
\rho(\tau)= Z^{-1}(\tau) e^{-\beta H(\lambda(\tau))}
\label{rtau}
\end{equation}
denotes the canonical state of a system at the initial temperature and the final parameter value $\lambda(\tau)$. 
Note that the coefficients obey the symmetry relation $c_{k,l} = c^*_{l,k}$.
Because the coefficients $c_{k,l}$ are independent of the phases $\phi_k$, differentiating both sides of Eq.~(\ref{eJE1}) with respect to $\phi_j$ yields
\begin{equation}
\mathop{\sum_l}_{l \neq j} \sin (\phi_j-\phi_l + \mu_{j,l}) |c_{j,l}| =0\:,
\label{eJE2} 
\end{equation}
where $\mu_{k,l} = -\mu_{l,k}$ is the phase of the coefficient $c_{k,l}$ whose absolute value is symmetric, $|c_{k,l}| =|c_{l,k}|$.
Equation (\ref{eJE2}) must hold for all choices of the phases $\phi_k$. This implies that all nondiagonal coefficients $c_{j,l}$, $j\neq l$, vanish.  Hence we find with Eq.~(\ref{ckl}) that
\begin{equation}
\langle \varphi_k | \int_X dx\, M_x M^\dagger_x | \varphi_l \rangle \langle \varphi_l| \rho(\tau) |\varphi_k\rangle = 0  
\label{eJE3}
\end{equation}
must hold for all $|\varphi_k\rangle$ and $|\varphi_l\rangle$ with $k \neq l$, 
taken from any complete orthonormal basis set $\{|\varphi_k \rangle \}$.
Because the density matrix $\rho(\tau)$ cannot be diagonal with respect to an arbitrary basis, Eq.~(\ref{eJE3}) implies that the nondiagonal matrix elements of $\int_X dx\, M_xM^\dagger_x$ must vanish with respect to any basis. This implies that $\int_X dx\, M_xM^\dagger_x$ is proportional to the identity, i.e.,
\begin{equation}
\int_X dx\, M_xM^\dagger_x = c \mathbbm{1}\:.
\end{equation}
Putting this expression via (\ref{ckl}) in Eq.~(\ref{eJE1}) yields $c=1$.
This concludes the proof that 
\begin{equation}
\int_X dx\, M_xM^\dagger_x =  \mathbbm{1}
 \label{nJE}
\end{equation}
is also a necessary condition for the validity of the Jarzynski equality in the presence of a universal measurement interrupting the force protocol.

\section{Necessary condition for the Crooks relation}
\label{A2}

The condition (\ref{CRxc1}) can be written as
\begin{equation}
\Tr\, U^\dagger_{t_1,0}(\Lambda) Y U_{t_1,0}(\Lambda) e^{-iu H(\lambda(0))} e^{-\beta H(\lambda(0))} =0\:,
\label{CRxc2}
\end{equation}
with
\begin{equation}
\begin{split}
Y &= \bar{\tilde{M}}_x U^\dagger_{\tau,t_1}(\Lambda) e^{iu H(\lambda(\tau))}  U_{\tau,t_1}(\Lambda) \bar{\tilde{M}}^\dagger_x\\
&\quad - M^\dagger_x U^\dagger_{\tau,t_1}(\Lambda) e^{iu H(\lambda(\tau))}  U_{\tau,t_1}(\Lambda) M_x  \:.
\end{split}
\label{Y}
\end{equation}
We suppose that the measurements during the forward and backward protocols are described by operators $M_x$ and $\tilde{M}_x$, respectively,  which are independent of (i) the protocol, (ii) the initial Hamiltonian, (iii) the final Hamiltonian, and (iv) the initial temperature of the system. It follows from (ii) and (iv) that the operator $Y$ must vanish, leading to the condition
\begin{equation}
\bar{\tilde{M}}_x Z \bar{\tilde{M}}^\dagger_x =  M^\dagger_x Z M_x\:,
\label{MZM}
\end{equation}
where 
\begin{equation}
Z = U^\dagger_{\tau,t_1}(\Lambda) e^{iu H(\lambda(\tau))}  U_{\tau,t_1}(\Lambda)\:.
\label{Z}
\end{equation}
One can obtain any linear operator by constructing linear combinations of the operators $Z$ for different protocols and final Hamiltonians. Therefore, the condition (\ref{MZM}) must hold for any linear operator $Z$, in particular also for $Z=\mathbbm{1}$, yielding
\begin{equation}
\bar{\tilde{M}}_x  \bar{\tilde{M}}^\dagger_x =  M^\dagger_x  M_x \equiv P_x^2\:,
\label{MZM1}
\end{equation} 
where $P_x$ is the positive semidefinite square root of $M_x^\dagger M_x$. Hence, using the polar representation one can write the measurement operators as
\begin{align}
\bar{\tilde{M}}_x &=P_x V_x \:,\quad &\bar{\tilde{M}}^\dagger_x &=  V^\dagger_xP_x\:, \label{MVP}\\
M_x &= U_xP_x \:, & M^\dagger_x &=P_x U^\dagger_x \:,
\label{MUP}
\end{align}  
with unitary operators $U_x$ and $V_x$. For given measurement operators $M_x$ and $\tilde{M}_x$, they are uniquely defined on the complement of the kernel of $P_x$ and arbitrary on the kernel itself. 
Putting the polar representations (\ref{MVP}) and (\ref{MUP}) in the condition (\ref{MZM}), we find
\begin{equation}
P_xV_x Z V^\dagger_x P_x = P_x U^\dagger_x Z U_x P_x\:.
\label{PVZVP}
\end{equation}
On the restriction to the complement of the kernel of $P_x$, this simplifies to
\begin{equation}
V_x Z V^\dagger_x = U^\dagger_x Z U_x\:.
\label{VZV}
\end{equation}
Because  $U_x$ and $V_x$ are undetermined on the kernel of $P_x$, we may require that they agree there with each other such that Eq.~(\ref{VZV}) holds on the whole Hilbert space. As this equation is required to be satisfied for all operators $Z$, it follows that $V_x= U^\dagger_x$. This leads immediately to Eq.~(\ref{MthM}) and hence concludes the proof of the necessity of  the particular choice of the measurement operators $\tilde{M}_x = \theta M^\dagger_x \theta^\dagger$ for the backward protocol in terms of the forward measurement operators $M_x$ for selective universal measurements.


\begin{thebibliography}{99}
\bibitem{fn1} Thermal equilibrium states other than canonical ones also lead to fluctuation relations involving changes of other thermodynamic potentials such as entropy changes for microcanonical initial states \cite{CvBK,THM,TMYH} or grand potential changes for grand-canonical initial states \cite{YKT}. Here we restrict ourselves to the original formulations of the transient fluctuation theorems for canonical initial states.
\end{thebibliography}
\end{document}